\documentclass[prd,preprint,superscriptaddress,amsmath,amssymb,nofootinbib]{revtex4}
\usepackage{graphicx}
\usepackage{dcolumn}
\usepackage{bm}
\usepackage{amssymb}
\usepackage{amsmath}
\usepackage{epsfig}    
\usepackage{color}
\usepackage{slashed}
\usepackage{hhline}
\usepackage{youngtab}
\usepackage[colorlinks = true,
            linkcolor = blue,
            urlcolor  = blue,
            citecolor = blue,
            anchorcolor = blue]{hyperref}
\usepackage{multirow}

\def\be{\begin{equation}}
\def\ee{\end{equation}}
\newcommand{\bea}{\begin{eqnarray}}
\newcommand{\eea}{\end{eqnarray}}
\newcommand{\nn}{\nonumber}

\begin{document}



\title{A hidden gauged $U(1)$ addressing radiative neutrino mass,\\ dark matter, $(g-2)_{\mu}$, and $H_0$ tension}

\author{Ujjal Kumar Dey}
\email{ujjal@iiserbpr.ac.in}
\affiliation{Department of Physical Sciences, Indian Institute of Science Education and Research Berhampur, Transit Campus, Government ITI, Berhampur 760010, Odisha, India}

\author{Hiroshi Okada}
\email{hiroshi.okada@apctp.org}
\affiliation{Asia Pacific Center for Theoretical Physics (APCTP) - Headquarters San 31, Hyoja-dong, Nam-gu, Pohang 790-784, Korea}
\affiliation{Department of Physics, Pohang University of Science and Technology, Pohang 37673, Republic of Korea}

\date{\today}

\begin{abstract}
We propose a radiative seesaw model with a hidden $U(1)$ gauge symmetry. In order to have anomaly cancellations, we need to introduce several new fermions that contribute to muon anomalous magnetic dipole moment $(g-2)_\mu$ as well as explain the neutrino oscillation data. We also consider a fermionic dark matter candidate that correlates with $(g-2)_\mu$ and neutrino mass matrix at the same time. We show allowed regions in our input parameters satisfying several constraints. Finally, we also briefly discuss the possibility of the resolution of the Hubble tension via neutrino self-interactions mediated by a lighter hidden gauge boson without spoiling our model.  
\end{abstract}

\maketitle

\section{Introduction}
%
Explaining nonzero neutrino masses is one of the important issues to be resolved beyond the standard model (SM) of particle physics.
In addition, there are certain experimental observations, namely existence of dark matter (DM), anomalous magnetic dipole moment of muon $(g-2)_\mu$ etc., which can also not be explained by the SM. In the standard paradigm of cosmology, i.e, the $\Lambda$CDM model there exists a statistically significant discrepancy in the measurements of the Hubble constant $H_0$ which is crucial in the estimation of the expansion rate of the Universe, see~\cite{DiValentino:2021izs} for a review. This is famously known as Hubble tension. 
Even though there exist a lot of mechanisms to generate the non-vanishing neutrino masses, a radiative seesaw scenario is attractive due potentially to involving the DM candidate. For most of the cases, an additional symmetry is imposed to maintain the neutrino loop as well as assuring the stability of DM.
If one introduces a hidden $U(1)$ gauge symmetry to realize a radiative neutrino mass model, there is a possibility that  Hubble tension can also be explained~\cite{Berbig:2020wve}. In addition, the DM problem can be addressed by the lightest Majorana neutrino present in the model, the DM stability is ensured by a remnant symmetry such as $\mathbb{Z}_2$ after spontaneous breaking of the hidden $U(1)$ symmetry.
In this paper, we study a radiative seesaw model introducing a hidden $U(1)$ gauge symmetry. In order to have the gauge anomaly cancellations among fermions, several fields are introduced in which a DM candidate naturally emerges. In addition, we can explain sizable $(g-2)_\mu$ as a bonus of this model due to appropriate charge assignments. A new gauge boson arising from this model can, in principle, give rise to a self-interaction of neutrinos which can potentially explain the $H_0$ tension as well.
This paper is organized as follows. In Section~II, we present our model formulating the renormalizable Lagrangian, scalar sector, neutrino sector, gauge sector, dark matter, lepton flavor violations (LFVs), and $(g-2)_\mu$.
In Section~III, we carry out numerical analysis showing our allowed input parameters for normal and inverted hierarchies of the neutrino masses as well as $(g-2)_\mu$ and LFVs. In Section~IV, we discuss the possibility of explaining the $H_0$ tension in this model. Finally we summarise and conclude in Section~V.

\section{Model setup and constraints}
\label{sec:model}
\begin{table}[t!]
\begin{tabular}{|c||c|c|c|c||c|c|c|c|}\hline\hline  
~&~ $U_R,\ U_L$ ~&~ $D_R,\ D_L$ ~&~ $E_R,\ E_L$ ~&~ $N_R,\ N_L$ ~&~ $\eta$ ~&~ $\chi$ ~&~ $\varphi$ ~&~ $\varphi'$
~ \\\hline\hline 
$SU(3)_C$ & $\bm{3}$  & $\bm{3}$ & $\bm{1}$ & $\bm{1}$ & $\bm{1}$ & $\bm{1}$ & $\bm{1}$ & $\bm{1}$ \\\hline 
$SU(2)_L$ & $\bm{1}$  & $\bm{1}$  & $\bm{1}$  & $\bm{1}$  & $\bm{2}$  & $\bm{1}$  & $\bm{1}$ & $\bm{1}$   \\\hline 
$U(1)_Y$   & $\frac23$ & $-\frac13$ & $-1$ & $0$  & $\frac12$ & $0$  & $0$  & $0$  \\\hline
$U(1)_H$   & $4,\ 1$ & $-4,\ -1$ & $-4,\ -1$   & $4,\ 1$  & $4$  & $1$  & $3$  & $2$ \\\hline
\end{tabular}
\caption{ 
Charge assignments of the our fields
under $SU(3)_C\times SU(2)_L\times U(1)_Y\times U(1)_{H}$, where all the SM fields are zero charges under the $U(1)_H$ symmetry. We suppose that all the exotic fermions have three families.}
\label{tab:1}
\end{table}
In this section we review our model.
We introduce isospin singlet exotic quarks $U_{R(L)}$ and $D_{R(L)}$, and exotic leptons $E_{R(L)}$ and $N_{R(L)}$, where we impose nonzero charges under a hidden $U(1)$ gauge symmetry, $U(1)_H$.
In Table~\ref{tab:1}, we depict the relevant charges of the fields present in the model.
The $U(1)_H$ chiral anomalies are canceled among these fields~\cite{Cai:2018upp}, and we need at least two families of $N_{R(L)}$ in order to reproduce the neutrino oscillation data.  Thus, all the exotic fermions have to have two or more than two families in order to realize the radiative seesaw model.
Here, we fix three families for our setup.
As for the bosonic sector, we introduce two types of inert scalars, namely, $\eta$ and $\chi$,
where $\eta$ is an isospin doublet and $\chi$ is an isospin singlet. Furthermore, we introduce two singlets $\varphi$ and $\varphi'$ both of which have nonzero VEVs that contribute to the spontaneous breaking of $U(1)_H$. 
The new scalar contents and their charge assignments are also summarized in Table~\ref{tab:1}.
The SM Higgs is denoted by $H$ and its VEV is defined by $\langle H\rangle\equiv [0,v/\sqrt2]^T$.
Then the valid renormalizable Yukawa Lagrangian and  Higgs potential under these symmetries are given by,
\begin{align}
-{\cal L}_Y
&=  y_{\ell_{ii}} \bar L_{L_i} H e_{R_i}  + f_{ai} \bar E_{R_a}\eta^* L_{L_i}+ g_{ia} \bar e_{R_i} E_{L_a}\chi\nn\\
&~~+ y_{\eta_{ia}} \bar L_{L_i} \tilde \eta N_{R_a}  + y_{E_{aa}}\varphi \bar E_{L_a} E_{R_a}
+  y_{{N}_{aa}} \bar N_{L_a} N_{R_a} \varphi^*
+  y_{{N'}_{ab}} \bar N^C_{L_a} N_{L_b} \varphi' 
+ {\rm h.c.}, \label{Eq:yuk} \\
V&= \sum_{\phi}^{H,\eta,\varphi,\varphi',\chi}\left(\mu^2_\phi \phi^\dag \phi +  \lambda_\phi |\phi^\dag \phi|^2 \right)
+  \sum_{\phi\neq \phi'}^{H,\eta,\varphi,\varphi',\chi} \lambda_{\phi\phi'} |\phi|^2 |\phi'|^2
+ \lambda'_{H\eta} (H^\dagger \eta)(\eta^\dagger H)
\nn\\&
~~+ \mu_0 \chi \varphi^*\varphi'+ \mu_1 \chi^2 \varphi'^* 
+  \lambda_0 \chi^* \varphi^* \varphi'^2 +  \lambda_1 (H^\dag \eta) \varphi\chi 
+ {\rm h.c.}, \label{Eq:pot}
\end{align}
where $\tilde \eta \equiv i\sigma_2\eta^*$, $\sigma_2$ being the second Pauli matrix, $\lambda_{\phi\phi'}\equiv \lambda_{\phi'\phi}$,
lower indices $(a,b,i)=1, 2, 3$ are the number of families. The
Yukawa couplings $y_{\ell},\ y_E,\ y_N$ can be taken to be diagonal by field-phase redefinitions without loss of generality.

\subsection{Scalar sector}
\label{sbsc:scalarSec}
%
At first, we define each scalar as follows,
\begin{align}
\label{eq:scalars}
&H =\left[\begin{array}{c}
w^+\\
\frac{v + h +i z}{\sqrt2}
\end{array}\right],\quad 
\eta =\left[\begin{array}{c}
\eta^+\\
\frac{ \eta_R + i \eta_I}{\sqrt2}
\end{array}\right],\nn\\
&\varphi\equiv \frac{v_\varphi+r+z_\varphi}{\sqrt2},\ 
\varphi' \equiv \frac{v_{\varphi'}+r'+z_{\varphi'}}{\sqrt2},\ 
\chi\equiv \frac{\chi_R + \chi_I}{\sqrt2},
\end{align}
where $w^\pm$ is absorbed by the SM singly-charged gauge boson $W^\pm$, and one degree of freedom in the CP-odd scalar sector $z$ and the Nambu-Goldstone boson $G$ after diagonalizing $z_{\varphi}$ and $z_{\varphi'}$ are respectively eaten by the neutral SM gauge boson $Z$ and the hidden gauge boson $Z'$.~\footnote{In our model, there is one physical Goldston boson and its phenomenology was discussed in ref.~\cite{Cai:2018upp}.} 
%
%
Here, we define the relevant mass eigenstates and their mixing matrices $O_{R(I)}$ for inert bosons $\chi$ and $\eta$.
Due to non-trivial terms $\mu$ and $\lambda_0$, they mix with each other.
\begin{align}
& \left( \begin{array}{c} \chi_R \\ \eta_R \end{array} \right) = \left( \begin{array}{cc} \cos \theta_R & - \sin \theta_R \\ \sin  \theta_R & \cos  \theta_R \end{array} \right) 
\left( \begin{array}{c} H_1 \\ H_2 \end{array} \right),  \\
& \left( \begin{array}{c} \chi_I \\ \eta_I \end{array} \right) = \left( \begin{array}{cc} \cos \theta_I & - \sin \theta_I \\ \sin  \theta_I & \cos  \theta_I \end{array} \right) 
\left( \begin{array}{c} A_1 \\ A_2 \end{array} \right),  
\label{Eq:scalar-mixing}
\end{align}
where $O_R(O_I)$ is the matrix parameterized by $\theta_R(\theta_I)$ above.
Notice here that we neglect the mixing among $h,r,r'$ for simplicity that would be favored by experiments.

\subsection{Neutrino sector}
\label{sbsc:nuSec}
After spontaneous symmetry breaking,
we have a neutral fermion mass matrix of 6$\times$6 in the basis of $\left(N_R^{C}, N_L \right)$. This is given by,
\begin{align}
M_N
&=
\left[\begin{array}{cc}
0 & m_D \\ 
m_D & M_{N_L} \\ 
\end{array}\right],
\end{align}
where $m_D\equiv y_N v_{\varphi}/\sqrt2$, $M_{N_L}\equiv y_{N'} v_{\varphi'}/\sqrt2$.   
$M_N$ is diagonalized by a unitary matrix $V_N$ as $D_N\equiv V_N^T M_N V_N$, $N_{R_i}\equiv \sum_{a=1}^6 V_{N_{ia}} \psi_{R_a}$
and $N^C_{L_i}\equiv \sum_{a=1}^6V_{N_{i+3,a}}\psi_{R_a}$ ($i=1,2,3$).
The entries of the diagonal matrix $D_N$ represent the mass eigenvalues and $\psi_{R_a}$ are the mass eigenstates.
The relevant Lagrangian in terms of mass eigenstate is found to be
\begin{align}
-{\cal L} = 
\frac{1}{\sqrt2} 
\bar \nu_{L_i} y_{\eta_{ia}} V_{N_{ab}} \psi_{R_{b}} (s_R H_1+c_R H_2)
-\frac{i}{\sqrt2} 
\bar \nu_{L_i} y_{\eta_{ia}} V_{N_{ab}} \psi_{R_{b}} (s_I A_1+c_I A_2) +{\rm h.c.},
\end{align}
where we have used short-hand notations $s_{R(I)}$ $c_{R(I)}$ for $\sin\theta_{R(I)}$ and $\cos\theta_{R(I)}$, respectively. 
Then, the neutrino mass matrix is given by~\cite{Cai:2018upp}
\begin{align}
(m_{\nu})_{ij} &= 
\frac{(y_\eta V_N)_{ia} D_{N_a} (y_\eta V_N)^T_{aj} }{2(4\pi)^2} 
\left[-2 F_I^a+ F_{II}^a \right],\\
 F_{I}^a&=\int[dx]_3 
\ln\left[\frac{x D^2_{N_a} + y m^2_{H_1}+ z m^2_{H_2}}{x D^2_{N_a} + y m^2_{A_1}+ z m^2_{A_2}}\right],\\
 F_{II}^a&=\int[dx]_3 \left[
 \frac{c^2_R m^2_{H_1}+ s^2_R m^2_{H_2}}{x D^2_{N_a} + y m^2_{H_1}+ z m^2_{H_2}}
 -
 \frac{c^2_I m^2_{A_1}+ s^2_I m^2_{A_2}}{x D^2_{N_a} + y m^2_{A_1}+ z m^2_{A_2}}
 \right],
\end{align}
where $\int[dx]_3\equiv \int_0^1 dx\int_0^{1-x}dy|_{z=1-x-y}$.\\
The neutrino mass matrix is diagonalized by unitary matrix $U_{MNS}$; $D_\nu= U_{MNS}^T m_\nu U_{MNS}$, where $D_\nu\equiv {\rm diag}(m_1,m_2,m_3)$. 
Then, we parametrize $y_\eta$ in terms of neutrino experiments and some model parameters as follows~\cite{Nomura:2018ktz},
\begin{align}
\label{eq:yeta}
y_\eta = U_{MNS}^* \sqrt{D_\nu} O_{\rm mix}  R_N^{-1} \lesssim \sqrt{4\pi},
\end{align}
where the last inequality suggests the dimensionless couplings $y_\eta$ has to be perturbative.
Here $O_{\rm mix}$ is an arbitrary $3 \times 3$ orthogonal matrix where this matrix can be parametrized by three angles $\alpha,\beta,\gamma$ as a standard parametrization, and $R_N$ is a lower unit triangular matrix~\cite{Nomura:2016run}, which is uniquely decomposed to be $(V_N D_N V_N^T)_{3\times3}=(R_N R^T_N)_{3\times3}$, since it is symmetric.

\subsection{Gauge sector}
\label{sbsc:gaugeSec}
%
Since we have an additional $U(1)_H$ gauge symmetry, the most general $U(1)$ gauge Lagrangian including the kinetic mixing with the SM $U(1)_Y$ is given as follows,
\begin{equation}
\mathcal{L}_{\text{gauge}}= -\frac{1}{4}B_{\mu\nu}B^{\mu\nu} - \frac{1}{4}B^\prime_{\mu\nu}B^{\prime\mu\nu} - \frac{1}{2}\xi B_{\mu\nu}B^{\prime \mu\nu} ,
\label{Lgauge}
\end{equation} 
where $B_{\mu\nu}$ and $B^\prime_{\mu\nu}$ are the field strength tensors of $U(1)_Y$ and $U(1)_H$ gauge symmetries, respectively.  We then diagonalize Eq.~\eqref{Lgauge} by the following transformation:
\begin{eqnarray}
\left(\begin{array}{c}
\tilde{B}^\prime_\mu\\
\tilde{B}_\mu\\
\end{array}\right)=\left(\begin{array}{cc}
\sqrt{1 - \xi^2} & 0 \\
\xi & 1 \\
\end{array}\right)\left(\begin{array}{c}
B^\prime_\mu\\
B_\mu\\
\end{array}\right)
\label{Gl2R}
\end{eqnarray}
where $\xi$ is a dimensionless quantity ($\xi \ll 1$) and we parameterize $\rho=-\xi/ \sqrt{1-\xi^2}$. Under the transformation Eq.~\eqref{Gl2R}, the gauge Lagrangian can be rewritten as follows,
\begin{eqnarray}
\mathcal{L}_{\text{gauge}}=-\frac{1}{4}\tilde{B}_{\mu\nu}\tilde{B}^{\mu\nu}-\frac{1}{4}\tilde{B}^\prime_{\mu\nu}\tilde{B}^{\prime\mu\nu}
\end{eqnarray}
where $\tilde{B}_{\mu\nu}=\partial_\mu \tilde{B}_\nu - \partial_\nu \tilde{B}_\mu$ and $\tilde{B}^\prime_{\mu\nu}=\partial_\mu \tilde{B}^\prime_\nu - \partial_\nu \tilde{B}^\prime_\mu$.
The kinetic term of the scalar fields with nonzero VEVs is 
\begin{eqnarray}
\mathcal{L}_{\text{kin}} = (D_{\mu}H)^\dagger (D^{\mu} H) 
+ (D_{\mu}\varphi)^\dagger (D^{\mu} \varphi) + (D_{\mu}\varphi')^\dagger (D^{\mu} \varphi').
\label{eq:scalar-kinetic}
\end{eqnarray}
The covariant derivatives of scalar fields are written by
\begin{eqnarray}
D_{\mu}H &=& \Big(\partial_\mu + ig_1\frac{\tau^a}{2}W_{\mu}^a + i\frac{g_2}{2}\tilde{B}_\mu
+ i\frac{g_H}{2}\rho\tilde{B'}_\mu\Big)H, \nonumber \\
D_{\mu}\varphi &=& \Big( \partial_\mu-3 ig_H \frac{\rho}{\xi}\tilde{B}^\prime_\mu \Big)\varphi, \nonumber \\
D_\mu \varphi' &=& \Big( \partial_\mu - 2i g_H \frac{\rho}{\xi}\tilde{B}^\prime_\mu \Big)\varphi',
\label{Covariant derivatives}
\end{eqnarray}
where $g_1$, $g_2$ and $g_H$ are respectively gauge couplings of $SU(2)_L$, $U(1)_Y$ and $U(1)_H$, $W^a$ are the $SU(2)_L$ gauge fields, and $\tau^a$ are the Pauli matrices. 

The masses of the gauge bosons come from Eqs.~\eqref{eq:scalar-kinetic} and \eqref{Covariant derivatives}. The mass matrix written in the basis of neutral gauge fields $(W_\mu^3, \tilde{B}_\mu, \tilde{B}^\prime_\mu)$ is 
\begin{eqnarray}
\mathcal{L}_{\text{gauge}}^{\text{mass}} = \frac{1}{2}
\left(\begin{array}{c}
W_\mu^3 \\
\tilde{B}_\mu\\
\tilde{B}^\prime_\mu
\end{array}\right)^T M^2_{\text{gauge}}
\left(\begin{array}{c}
W_\mu^3 \\
\tilde{B}_\mu\\
\tilde{B}^\prime_\mu
\end{array}\right) ,
\end{eqnarray} 
where
\begin{eqnarray}
M^2_{\text{gauge}}=\frac{1}{4}
\left(\begin{array}{ccc}
g_1^2 v^2 & -g_1g_2 v^2 & -g_1g_2 v^2 \rho \\
-g_1 g_2 v^2 & g_2^{2} v^2 & g_2^{2} v^2 \rho\\
-g_1 g_2 v^2 \rho & g_2^{2} v^2 \rho 
&g_2^{2} v^2 \rho 
+ 4g_H^2(9v^2_{\varphi}+4 v^2_{\varphi'})\frac{\rho^2}{\xi^2} \\
\end{array}\right).
\label{masss matrix gauge boson}
\end{eqnarray}
Here we parameterize 
\begin{align}
M_{Z^\prime}^2 = \frac{1}{4} g_2^{2} v^2 \rho^2 + g_H^2(9v^2_{\varphi}+4 v^2_{\varphi'})\frac{\rho^2}{\xi^2}. 
\label{eq:MZprimeSq}
\end{align}
We rotate the fields $(W_\mu^3,\tilde{B}_\mu)$ by Weinberg
angle $\theta_W$ to obtain the massless photon field $A_\mu$~\footnote{In general, Weinberg angle is defined to be the mixing between $W^3_\mu$ and $B_\mu$ in the SM, this can be realized by $\rho\to 0$. In fact, we expect $\rho$ be so small.}
\begin{eqnarray}
\left(\begin{array}{c}
W^3_\mu \\
\tilde{B}_\mu \\
\end{array}\right) =
\left(\begin{array}{cc}
\cos\theta_W & \sin\theta_W \\
-\sin\theta_W & \cos\theta_W\\
\end{array}\right)
\left(\begin{array}{c}
\tilde{Z}_\mu \\
A_\mu \\
\end{array}\right).
\end{eqnarray}
And the mass matrix for the massive neutral gauge bosons is given as follows:
\begin{eqnarray}
&& \mathcal{L}^{\text{mass}}_{\text{gauge}} = \frac{1}{2}
\left(\begin{array}{c}
\tilde{Z}_\mu \\
\tilde{B}^\prime_\mu \\
\end{array}\right)^T
\left(\begin{array}{cc}
M_{Z,SM}^2 & -\Delta^2 \\
-\Delta^2 & M_{Z'}^2
\end{array}\right)
\left(\begin{array}{c}
\tilde{Z}_\mu \\
\tilde{B}^\prime_\mu \\
\end{array}\right),
\label{masss matrix gauge boson 2}
\end{eqnarray}
where $\Delta^2 = \frac{1}{4} g_2 v^2 \rho \sqrt{g_1^2 +g_2^{2}}$ 
and $M_{Z,SM}^2 = \frac{1}{4} v^2 (g_1^2 + g_2^{2})$.
The physical masses are 
\begin{eqnarray}
m_Z^2 &=& \frac{1}{2}\Big[M_{Z,SM}^2 + M_Z^{\prime 2} + \sqrt{(M_{Z,SM}^2 - M_{Z^\prime}^2)^2 + 4\Delta^4} \Big],\nonumber \\
m_{Z^{\prime}}^2 &=& \frac{1}{2}\Big[M_{Z,SM}^2 + M_Z^{\prime 2} - \sqrt{(M_{Z,SM}^2 - M_{Z^\prime}^2)^2 + 4\Delta^4} \Big],
\label{eq:mzmzpMass}
\end{eqnarray}
where we expect $m_Z \gg m_{Z'}$ to address the Hubble tension well.
The mass matrix in Eq.~\eqref{masss matrix gauge boson 2} can be diagonalized by rotation matrix 
\begin{align}
\left(\begin{array}{c}
\tilde{Z}_\mu\\
\tilde{B}^\prime_\mu\\
\end{array}\right)&=
\left(\begin{array}{cc}
\cos\epsilon & \sin\epsilon\\
-\sin\epsilon & \cos\epsilon \\
\end{array}\right)
\left(\begin{array}{c}
Z_\mu\\
Z^\prime_\mu\\
\end{array}\right), \\
\tan 2\epsilon &= \frac{2\Delta^2}{M_{Z,SM}^2 - M_{Z^\prime}^2}
\approx 2\frac{g_2\rho}{\sqrt{g_1^2+g_2^2}} 
= 2\rho\sin\theta_W \approx \rho,
\end{align}
where $\sin\theta_W^2\approx 0.2229$, $Z_\mu$ and $Z^\prime_\mu$ are the two physical gauge bosons which respectively correspond to the SM $Z$ boson and extra gauge boson. 
The above relation suggests $\epsilon\approx\rho/2$ if $\epsilon$ is small enough.
The quantity $\rho$ is restricted by the absolute discrepancy between $m_Z$ and $M_{Z,SM}$.
It is given by
$|\Delta m_Z| \approx 
   M_{Z,SM}(\sqrt{1+g_2^4 \rho^2}-1)$ and $|\Delta m_Z|$ has to be within the experimental uncertainty as follows:
\begin{align}
|\Delta m_Z| \lesssim 0.0021 {\rm GeV}.
\end{align}
Finally, we obtain 
\begin{align}
\label{eq:epsLim}
\epsilon \lesssim 0.0312,
\end{align}
where we have used the following input values; $M_{Z,SM}=91.1876$ GeV, $g_2=0.33$.
However, the kinetic mixing receives several stringent constraints from astrophysical experiments in light mass scale of hidden gauge boson; 1 eV - 100 eV. These experiments suggest $\epsilon \ll {\cal O}(10^{-9})$~\cite{Jaeckel:2010ni} that is far from the constraint from electroweak precision test of $Z$.
Thus, we need to rely on mixing among neutral fermions via Yukawa terms. Therefore, $\eta$ has to have tiny VEV. 
In this case, the active neutrino mass is induced at tree level via inverse seesaw in general. But, we assume that this tree level neutrino mass is negligibly small compared to the one-loop contribution. Theoretically, it is easy to realize by controlling the parameter of VEV of $\eta$. Then, the corresponding mixing paramter is found to be $\tan\tilde\epsilon\sim f v_\eta/(\sqrt2 m_D)$~\cite{Berbig:2020wve}. 

\subsection{Dark matter}
\label{sbsc:dm}
Here, we suppose our DM to be the lightest Majorana fermion; $X_R\equiv\psi_{R_1}$. Here, we define the mass to be $M_X$.
Then, the relevant interaction Lagrangian is given by,
\begin{equation}
- \mathcal{L}_{DM} = \frac1{\sqrt2} \bar\nu_{L_i} G_{i1} X_R(s_R H_1+c_R H_2)
- \frac{i}{\sqrt2} \bar\nu_{L_i} G_{i1} X_R(s_I A_1+c_I A_2) + \bar\ell_{L_i} G_{i1} X_R \eta^-
+ {\rm h.c.}, \label{Eq:yuk_DM}
\end{equation}
where $G\equiv U_{MNS}^* \sqrt{D_\nu} O_{\rm mix}  R_N^{-1} V_N$.
Then, the thermally averaged annihilation cross section to explain the DM is $s$-wave dominant and given by 
\begin{align}
\langle\sigma v_{rel}\rangle\simeq
\frac{M_X^2 |G_{i1}G_{1j}^\dag|^2}{64\pi} &\left[\frac{4}{(M_X^2+m_{\eta^-}^2)^2} \right. \nonumber \\
&+
\left. \left(
\frac{s_R^2}{M_X^2+m_{H_1}^2} +\frac{c_R^2}{M_X^2+m_{H_2}^2} 
+
\frac{s_I^2}{M_X^2+m_{A_1}^2} +\frac{c_I^2}{M_X^2+m_{A_2}^2}
\right)^2
\right],
\end{align}
where we assume $m_{\eta^-}\approx m_{A_2}$ simply in order to evade the constraint from oblique parameters under the small mixing of $\theta_{I}$.
The resulting relic density is given by~\cite{Baek:2016kud},
\begin{align}
&\Omega h^2\approx 
\frac{1.07\times 10^9 x_f}{\sqrt{g_*(x_f)} M_{\rm Pl}\langle\sigma v_{rel}\rangle }
\approx\frac{2.2\times 10^{-10}}{\langle\sigma v_{rel}\rangle}
 \label{eq:relic-rl}~,
\end{align}
where the present relic density is $0.1199 \pm 0.0054$ at 2$\sigma$~\cite{Planck:2013pxb}, $g_*(x_f\approx 25)\approx100$ counts the degrees of freedom for relativistic particles, and $M_{\rm Pl}\approx 1.22\times 10^{19}$~GeV is the Planck mass.
In our numerical analysis below, however, we will use relaxed observable $0.11\le \Omega h^2\le 0.13$ which is at around 3$\sigma$.
%
%
\subsection{Lepton flavor violations and muon anomalous magnetic dipole moment $(g-2)_\mu$}
\label{sbsc:lfvgmin2}
Lepton flavor violations (LFV) as well as $(g-2)_\mu$ are arisen at one-loop level, and LFVs form is given by~\cite{Baek:2016kud}
\begin{align}
& {\rm BR}(\ell_i\to\ell_j\gamma)= \frac{48\pi^3\alpha_{\rm em} C_{ij} }{{\rm G_F^2} m_{\ell_i}^2}\left(|a_{R_{ij}}|^2+|a_{L_{ij}}|^2\right),\\
 a_{R_{ij}} &\approx
 - \sum_{a=1,2,3}  \frac{g_{ja} M_{E_a} f_{ai}}{(4\pi)^2}\nn\\
&\times 
\left[ s_R c_R (F[M_{E_a}, m_{H_1}]-F[M_{E_a}, m_{H_2}])
+
s_I c_I (F[M_{E_a}, m_{A_1}]-F[M_{E_a}, m_{A_2}]) \right],
\label{eq:agf}\\
\quad 
a_{L_{ij}} &\approx
- \sum_{a=1,2,3}  \frac{f^\dag_{ja} M_{E_a} g^\dag_{ai}}{(4\pi)^2}\nn\\
&\times 
\left[ s_R c_R (F[M_{E_a}, m_{H_1}]-F[M_{E_a}, m_{H_2}])
+
s_I c_I (F[M_{E_a}, m_{A_1}]-F[M_{E_a}, m_{A_2}]) \right],
\label{eq:afg}\\
F[m_1,m_2]&\approx\frac{m_1^2-m_2^2+ m_2^2 \ln\left[\frac{m_2^2}{m_1^2}\right]}{(m_1^2-m_2^2)^2},
\end{align}
where $M_{E}\equiv y_E v_\varphi/\sqrt2$ ${\rm G_F}\approx 1.17\times10^{-5}$[GeV]$^{-2}$ is the Fermi constant, $\alpha_{\rm em}\approx1/137$ is the fine structure constant, $C_{21}\approx1$, $C_{31}\approx 0.1784$, and $C_{32}\approx 0.1736$.
Experimental upper bounds are respectively given by ${\rm BR}(\mu\to e\gamma)\lesssim 4.2\times10^{-13}$, ${\rm BR}(\tau\to e\gamma)\lesssim 3.3\times10^{-8}$, and ${\rm BR}(\tau\to \mu\gamma)\lesssim 4.4\times10^{-8}$~\cite{MEG:2013oxv, MEG:2016leq}.
\\
New contribution to $(g-2)_\mu$ also arises from the same term as in LFVs,
and it is given by\footnote{For a comprehensive review on new physics models for the $(g-2)_\mu$ anomaly as well as lepton flavour violation, please see Ref.~\cite{Lindner:2016bgg}.}
\begin{align}
&\Delta a_\mu =- m_\mu [a_R+a_L]_{22}.
\label{eq:G2-ZP}
\end{align}
Combined  results with the previous BNL, suggests that $(g-2)_\mu$ deviates from the SM prediction by 4.2$\sigma$ level~\cite{Aoyama:2012wk, Aoyama:2019ryr, Czarnecki:2002nt, Gnendiger:2013pva, Davier:2017zfy, Keshavarzi:2018mgv, Colangelo:2018mtw, Hoferichter:2019mqg, Davier:2019can, Keshavarzi:2019abf, Kurz:2014wya, Melnikov:2003xd, Masjuan:2017tvw, Colangelo:2017fiz, Hoferichter:2018kwz, Gerardin:2019vio, Bijnens:2019ghy, Colangelo:2019uex, Blum:2019ugy, Colangelo:2014qya, Hagiwara:2011af},
\begin{align}
\Delta a_\mu = (25.1\pm5.9)\times10^{-10} ~.
\label{eq:damu}
\end{align}
%

 \section{Numerical analysis}
 \label{sec:numAna}
 In this numerical analysis, we randomly select absolute values of input parameters within the following ranges,
 \begin{subequations}
 \begin{align}
   \{|\alpha|,\ |\beta|,\ |\gamma|\}\ {\text{in}} \ O_{\rm mix} &\in [0, \pi],\\
     \{s_R,\ s_I,\ f,\ g\} &\in [0, 0.3],\\
     \{m_{H_1},\ m_{H_2},\ m_{A_1},\ m_{A_2}\} &\in [10^2, 10^5]~\text{GeV},\\
     \{m_D,\ M_{N_L}\} &\in [0.1, 10^5]~\text{GeV},
 \end{align}
 \end{subequations}
{where $f$ and $g$ are relevant Yukawa couplings as described in Eqs.~\eqref{eq:agf} and \eqref{eq:afg}.}
 Then, we also impose the perturbative limit for $y_\eta\lesssim \sqrt{4\pi}$ as discussed in Eq.~\eqref{eq:yeta} and the  neutrino observables are accommodated by \texttt{Nufit 5.1} as the best fit values in case of ``without SK atmospheric data"~\cite{nufit} vanishing Majorana phases, observed relic density, $0.11\lesssim \Omega h^2\lesssim 0.13$; and $(g-2)_\mu$ within 1$\sigma$, $19.2\times10^{-10}\lesssim \Delta a_\mu\lesssim 31.0\times10^{-10}$ in Eq.~(\ref{eq:damu}).

\subsection{Normal Hierarchy}
\begin{figure}[!htbp]
  \includegraphics[scale=0.25]{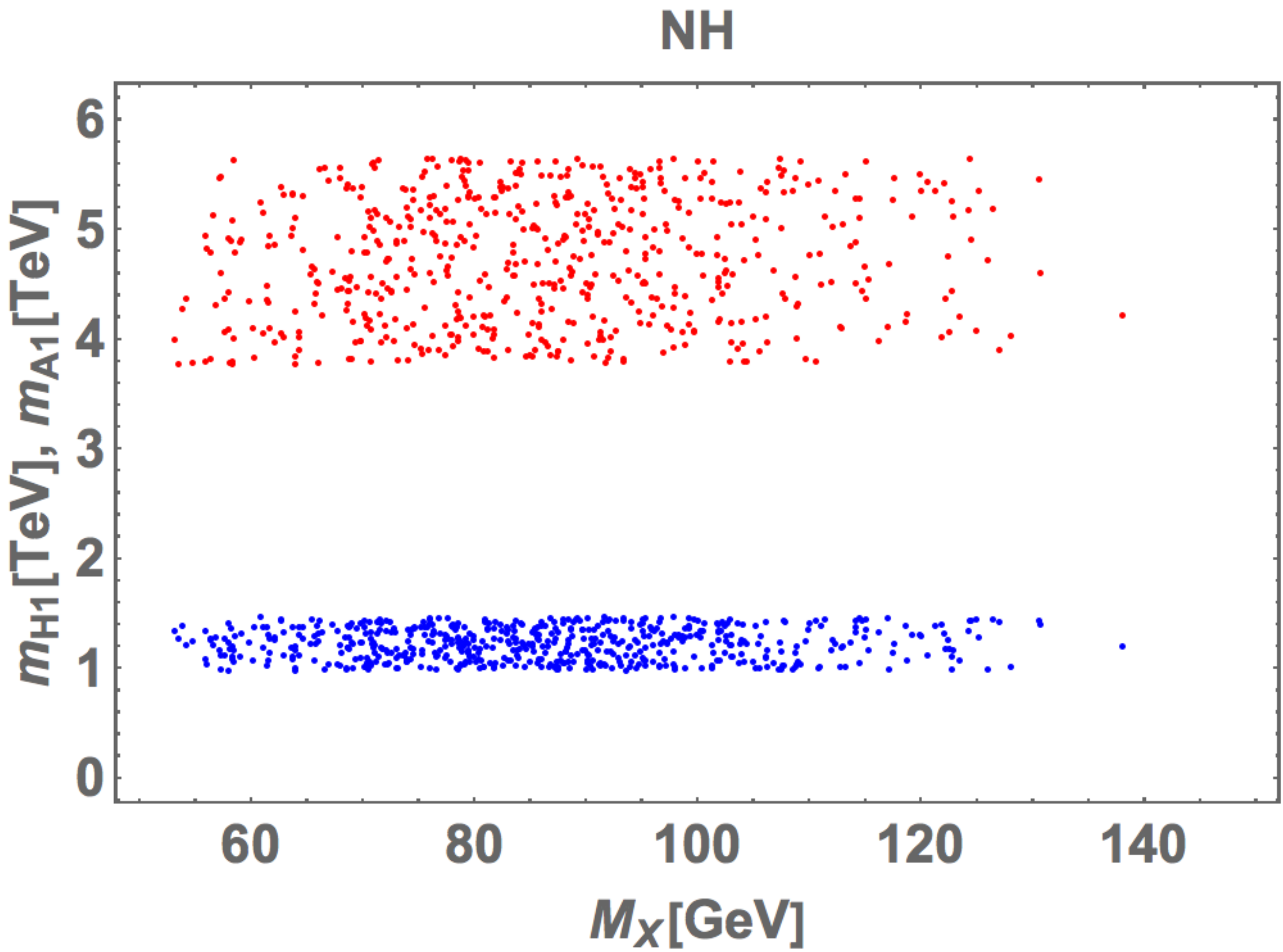}~~~~
  \includegraphics[scale=0.255]{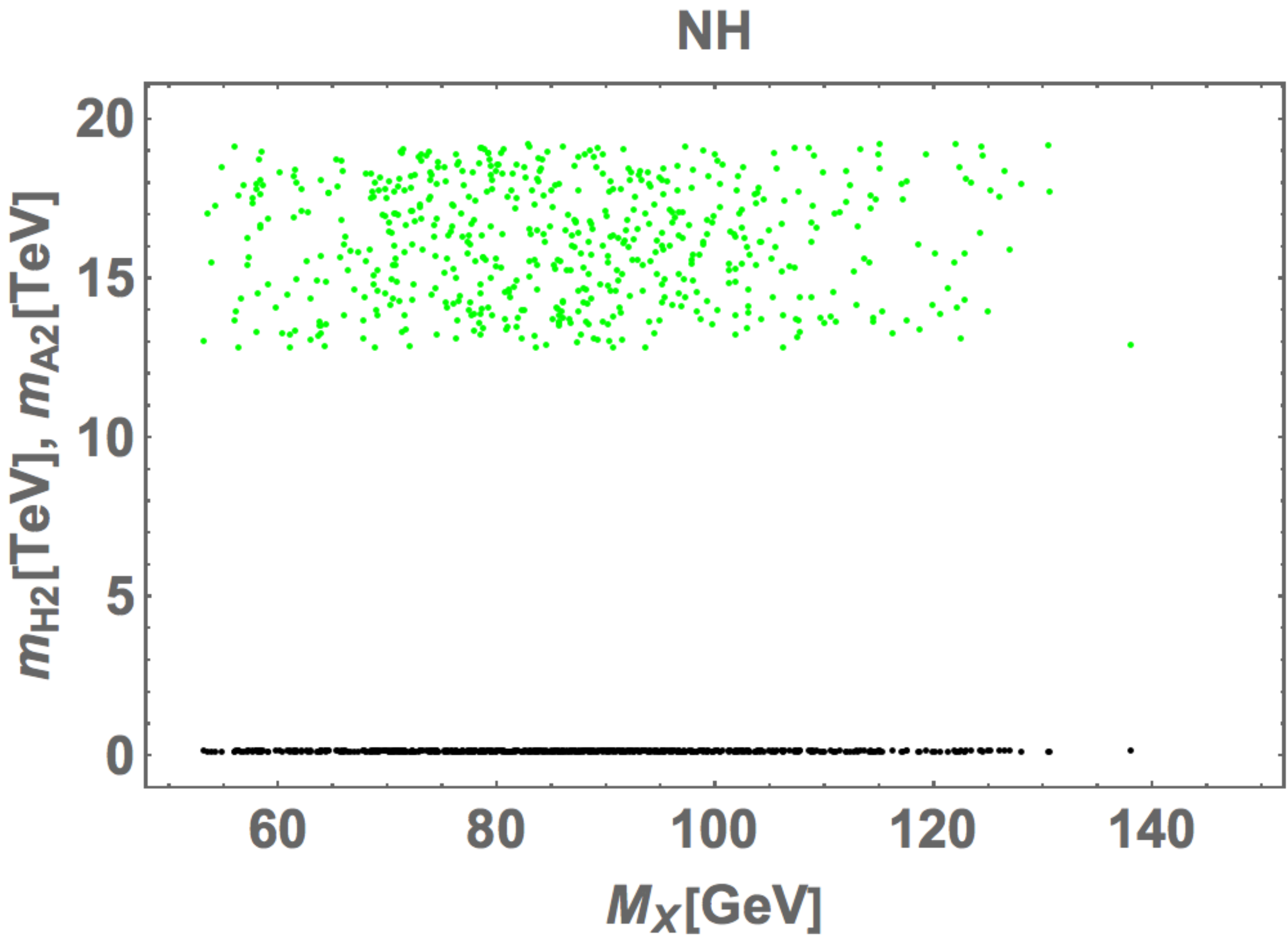}
 \caption{Allowed regions of $m_{H_1}$(blue) and $m_{A_1}$(red) in the left panel and the one of $m_{H_2}$(black) and $m_{A_2}$(green) in the right panel in terms of $M_X$ in GeV unit. Here, these points satisfy $0.11\le\Omega h^2\le0.13$.}
 \label{fig:dm}
\end{figure}
In Fig.~\ref{fig:dm}, we show the allowed regions of $m_{H_1}$(blue) and $m_{A_1}$(red)  in the left panel and the one of $m_{H_2}$(black) and $m_{A_2}$(green) in the right panel of Fig.~\ref{fig:dm} in terms of $M_X$. Here, these points satisfy $0.11\le\Omega h^2\le0.13$.
These figures suggest 50 GeV$\lesssim M_X\lesssim$140 GeV,
1000 GeV$\lesssim m_{H_1}\lesssim$1400 GeV, 3800 GeV$\lesssim m_{A_1}\lesssim$5600 GeV,
117 GeV$\lesssim m_{H_2}\lesssim$175 GeV, 12.8 TeV$\lesssim m_{A_2}\lesssim$19.23 TeV.

\begin{figure}[!htbp]
  \includegraphics[scale=0.25]{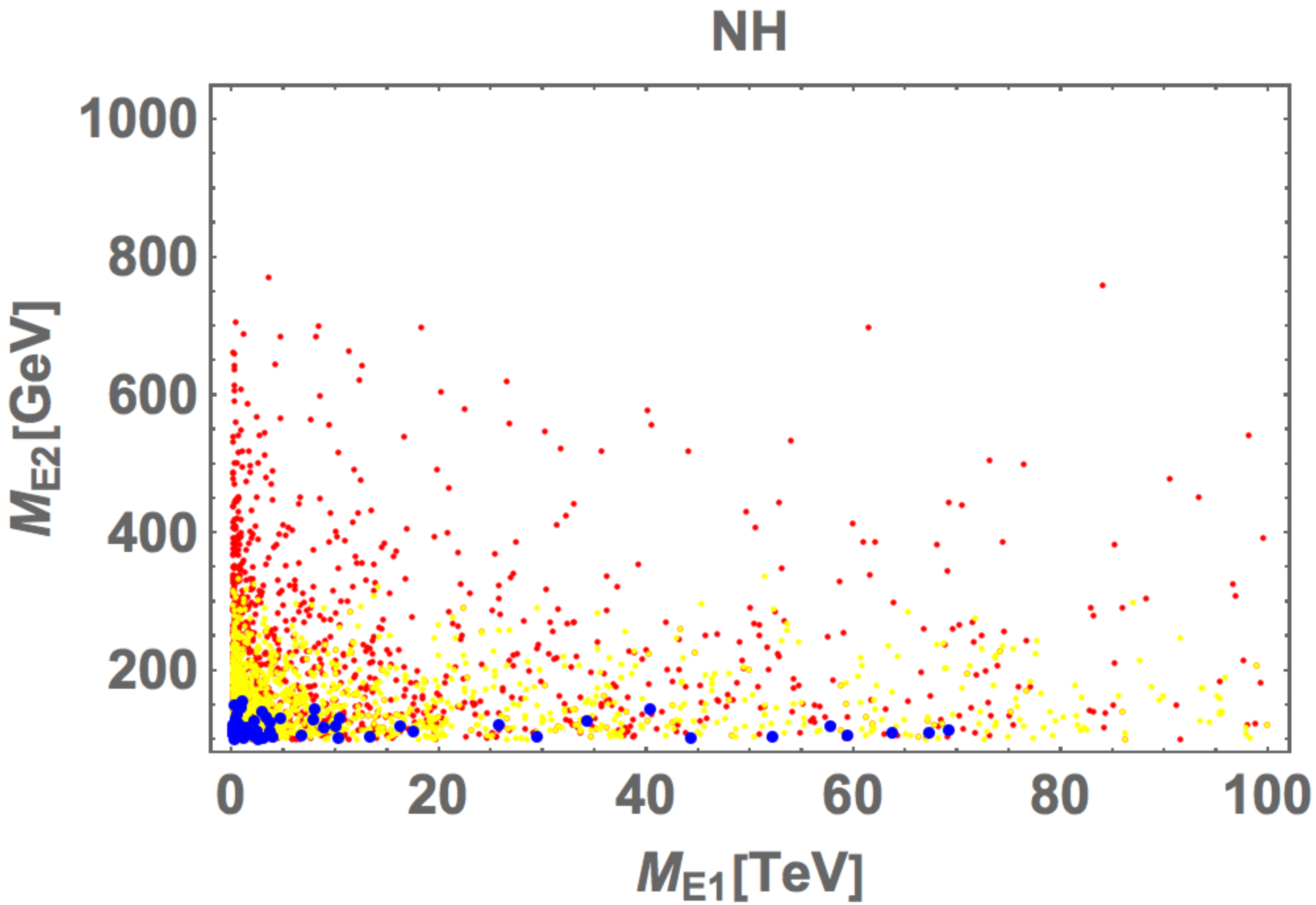}~~   \includegraphics[scale=0.36]{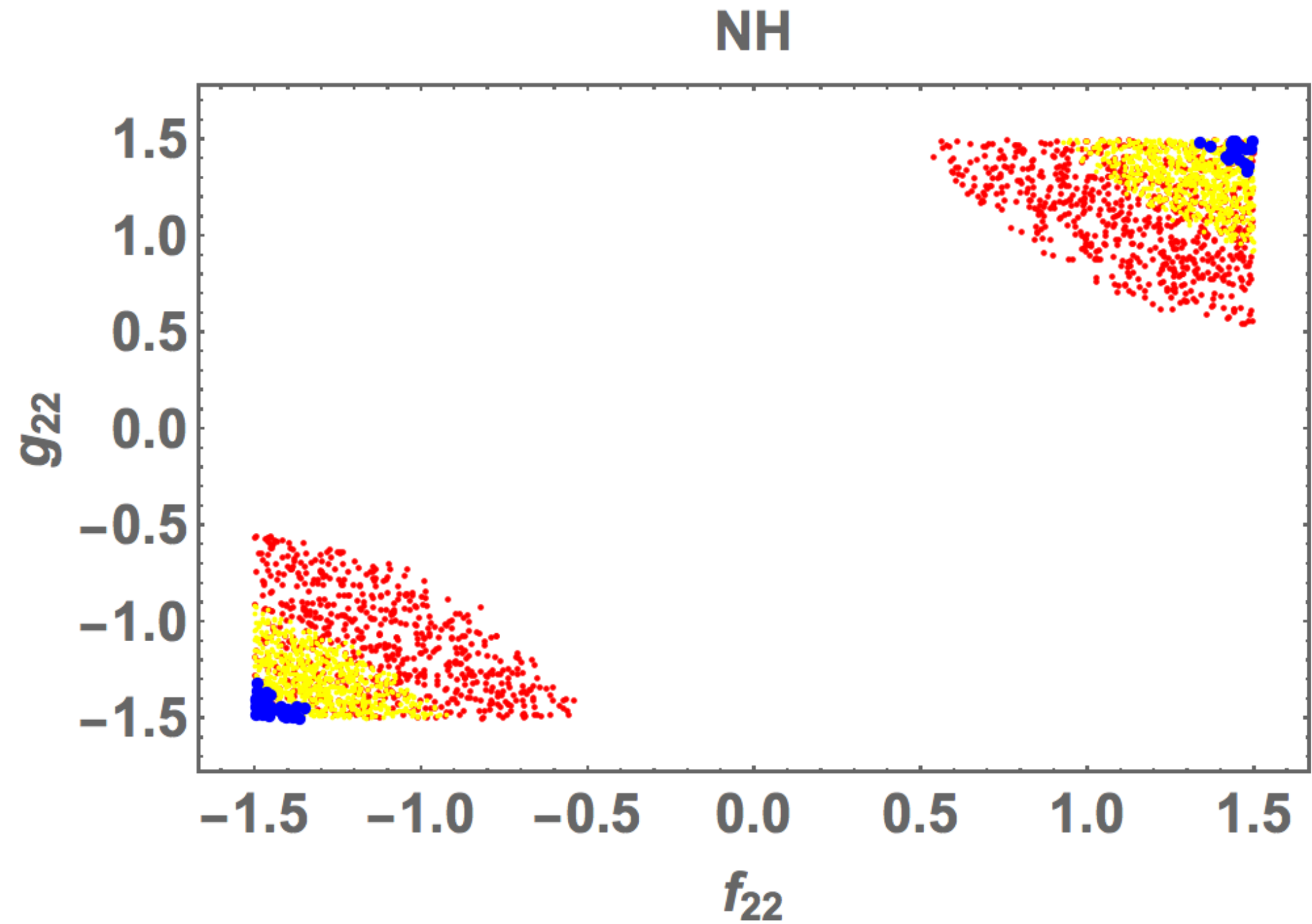}
 \caption{$(g-2)_\mu$ allowed region in $M_{E_1}$-$M_{E_2}$ plane (left panel); and $f_{22}$-$g_{22}$ plane (right panel). The blue color represents $(g-2)_\mu$ within 1$\sigma$ level, yellow one 2$\sigma$ level, and red one 3$\sigma$ level. Here, we have fixed $\Omega h^2=0.12299$.}
 \label{fig:admu}
\end{figure}
In Fig.~\ref{fig:admu}, we show the allowed regions to satisfy $(g-2)_\mu$
in terms of $M_{E_1}$ and $M_{E_2}$ in the left figure,
and $f_{22}$ and $g_{22}$ in the right figure. The blue color represents $(g-2)_\mu$ within 1$\sigma$ level, yellow one 2$\sigma$ level, and red one 3$\sigma$ level.
Here, we have subtracted a benchmark point from Fig.~\ref{fig:dm} so that we have fixed $\Omega h^2=0.12299$.
These figures suggest that  
100 GeV$\lesssim M_{E_1}\lesssim10^5$ GeV,
 and 100 GeV$\lesssim M_{E_2}\lesssim$800 GeV 
 and 0.5$\lesssim (|f_{22}|,\ |g_{22}|)\lesssim$1.5 within 3$\sigma$ level.

 \subsection{Inverted Hierarchy}
\begin{figure}[!htbp]
  \includegraphics[scale=0.29]{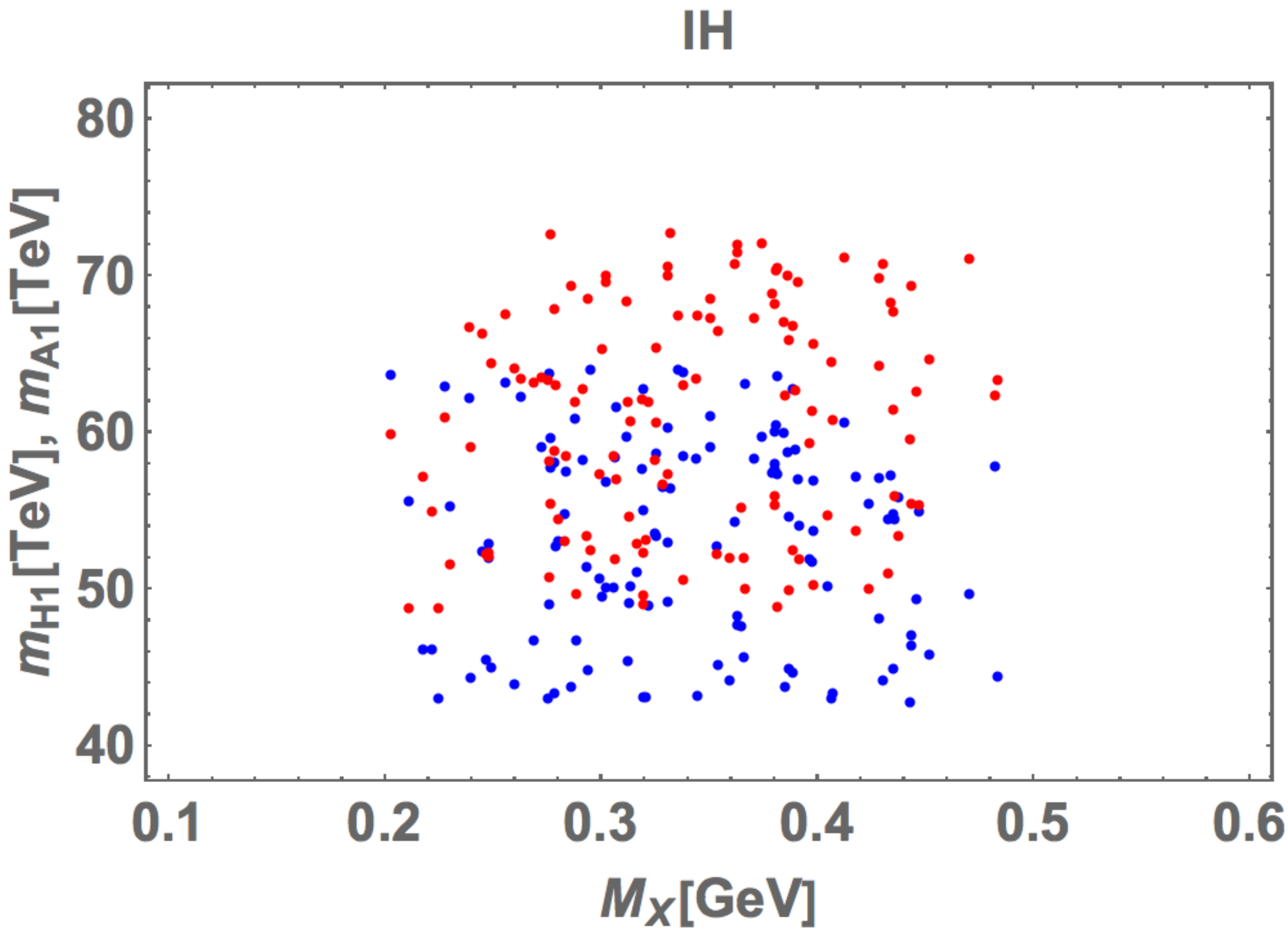}~~   \includegraphics[scale=0.3]{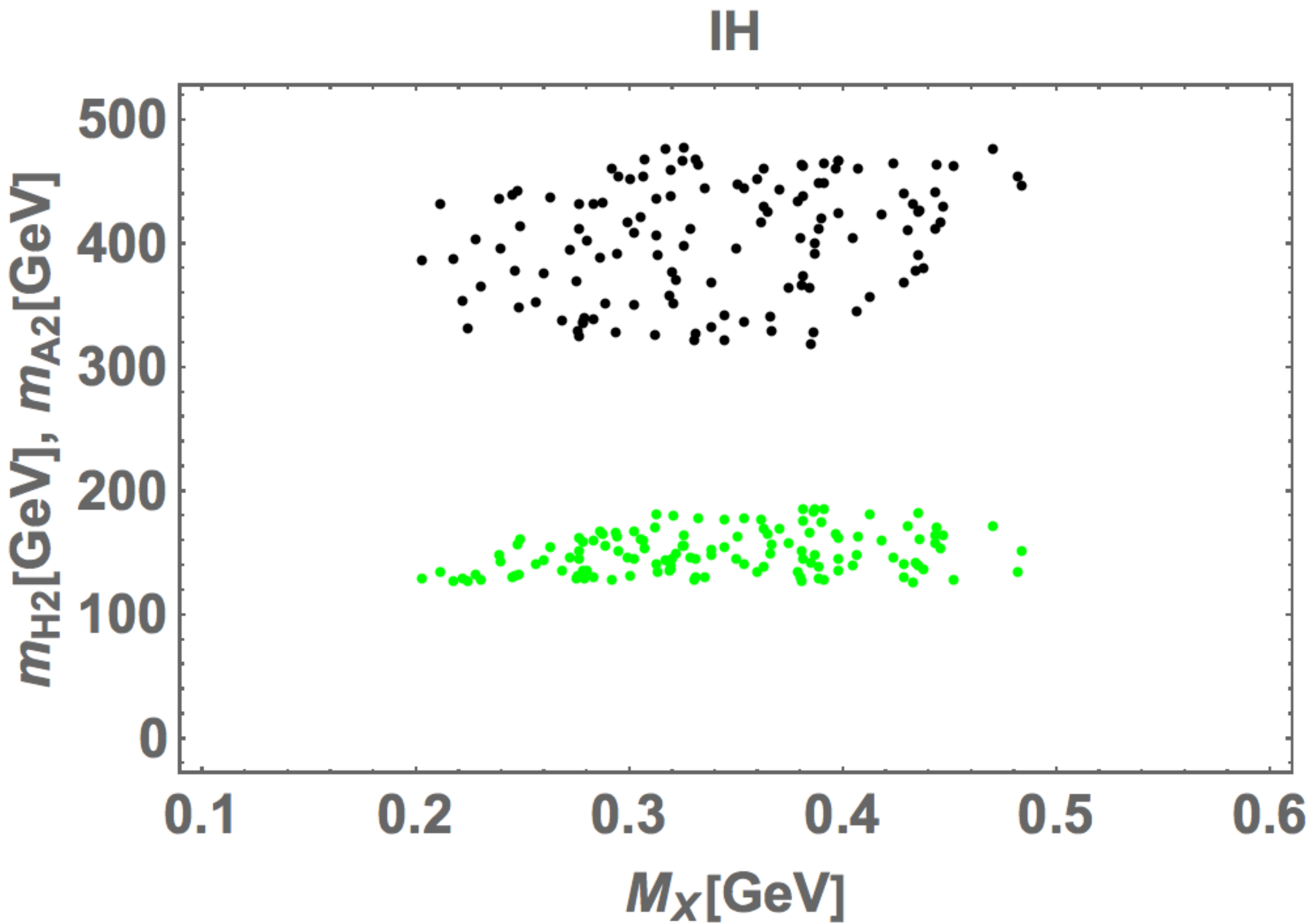}
 \caption{Allowed regions of $m_{H_1}$(blue) and $m_{A_1}$(red)  [left panel] and the one of $m_{H_2}$(black) and $m_{A_2}$(green) [right panel] in terms of $M_X$ in GeV unit. Here, these points satisfy $0.11\le\Omega h^2\le0.13$.}
 \label{fig:dm_ih}
\end{figure}
In Fig.~\ref{fig:dm_ih}, we show the allowed regions of $m_{H_1}$(blue) and $m_{A_1}$(red)  in the left panel and the one of $m_{H_2}$(black) and $m_{A_2}$(green) in the right panel in terms of $M_X$ in GeV unit. Here, these points satisfy $0.11\le\Omega h^2\le0.13$.
These figures suggest 0.2 GeV$\lesssim M_X\lesssim$0.5 GeV,
42 TeV$\lesssim m_{H_1}\lesssim$62 TeV, 48 TeV$\lesssim m_{A_1}\lesssim$72 TeV,
300 GeV$\lesssim m_{H_2}\lesssim$480 GeV, 120 GeV$\lesssim m_{A_2}\lesssim$180 GeV.

\begin{figure}[!htbp]
  \includegraphics[scale=0.29]{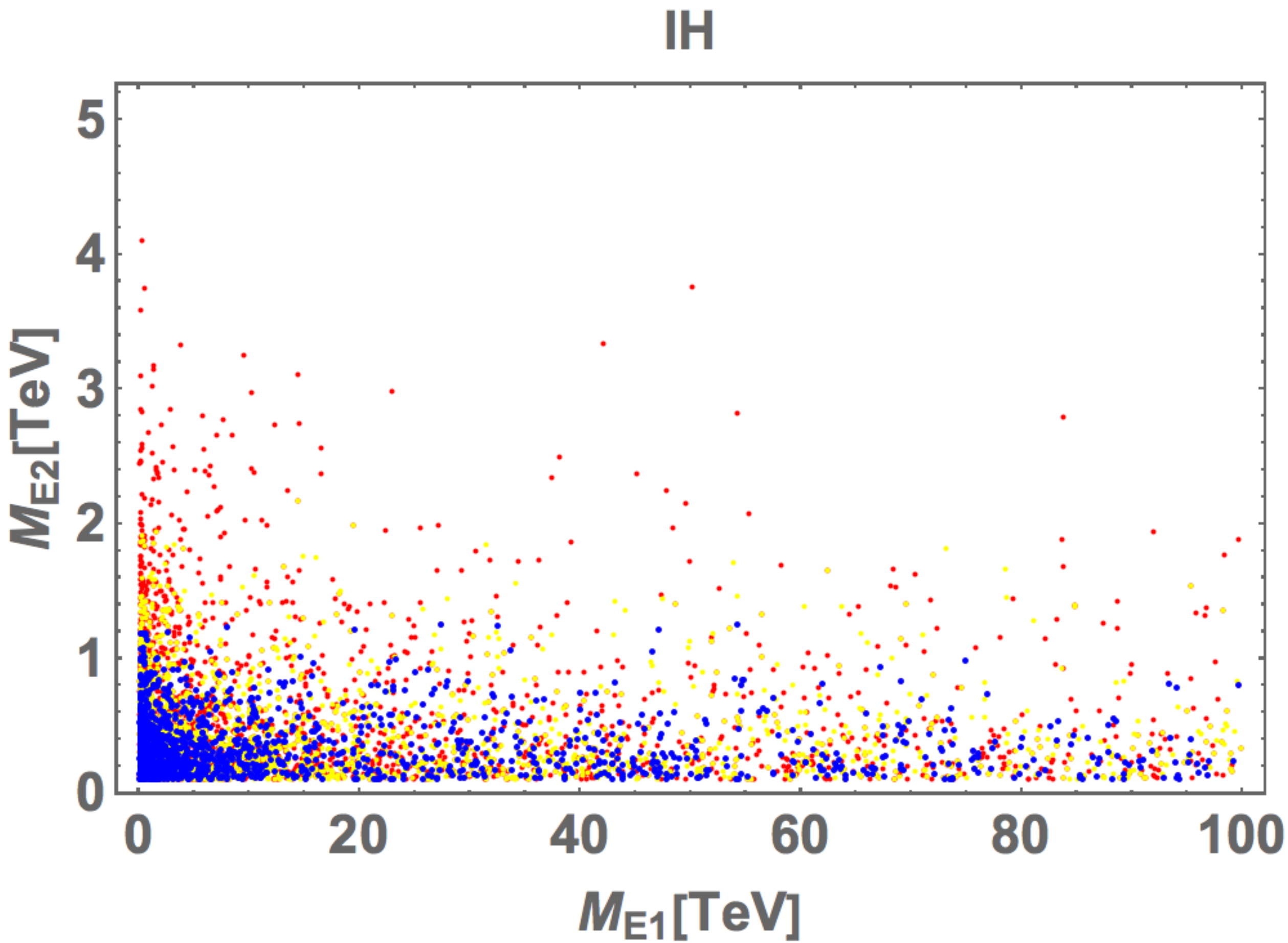}~~   \includegraphics[scale=0.3]{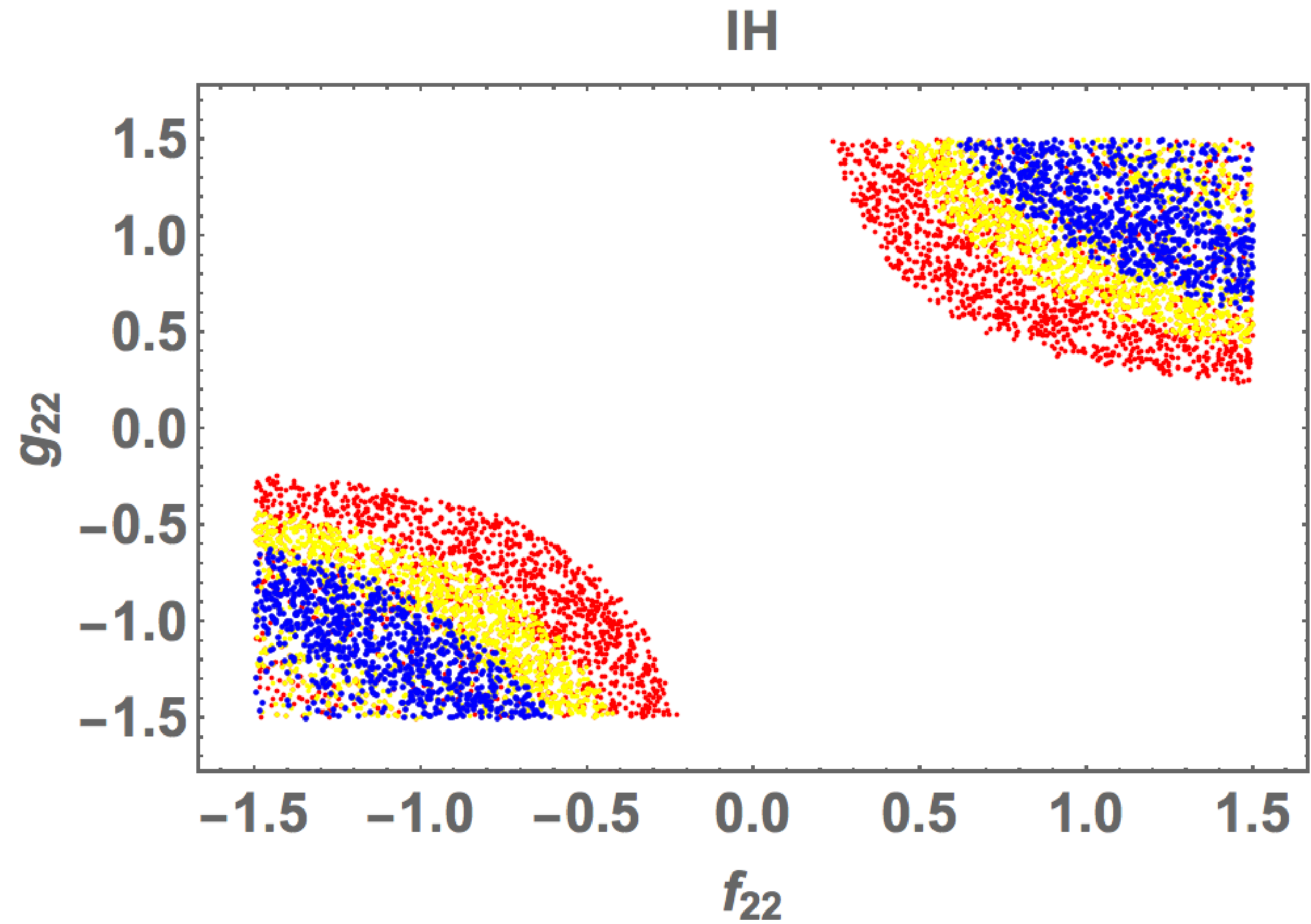}
 \caption{Muon $(g-2)$ allowed region in the $M_{E_1}$-$M_{E_2}$ plane (left panel), and in the $f_{22}$-$g_{22}$ plane (right panel). The blue color represents the muon $g-2$ within 1$\sigma$ level, yellow one 2$\sigma$ level, and red one 3$\sigma$ level. Here, we have fixed $\Omega h^2=0.118805$.}
 \label{fig:admu_ih}
\end{figure}
In Fig.~\ref{fig:admu_ih}, we show the allowed region that satisfies $(g-2)_\mu$ constraint in terms of $M_{E_1}$-$M_{E_2}$ plane (left) and $f_{22}$-$g_{22}$ plane (right). The blue color represents $(g-2)_\mu$ within 1$\sigma$ level, yellow one 2$\sigma$ level, and red one 3$\sigma$ level. Here, we have subtracted a benchmark point from Fig.~\ref{fig:dm_ih} and fixed $\Omega h^2=0.118805$.
These figures suggest that 100 GeV$\lesssim M_{E_1}\lesssim10^5$ GeV, and 100 GeV$\lesssim M_{E_2}\lesssim$4200 GeV and 0.2$\lesssim (|f_{22}|,\ |g_{22}|)\lesssim$1.5 within 3$\sigma$ level.

\section{Hubble tension}
\label{sec:hubble}
There has been a widely persisting and statistically significant ($4\sigma$-$6\sigma$) discrepancy in the measurements of the Hubble constant $H_0$ from late time and early time observations. There are attempts to tweak the standard $\Lambda$CDM to address this~\cite{Schoneberg:2019wmt, Knox:2019rjx}. Another attractive possible resolution comes from considering the scenario of self-interacting neutrinos~\cite{Blinov:2019gcj, He:2020zns, Berbig:2020wve, Lyu:2020lps, Das:2020xke, Brinckmann:2020bcn}. Such an effective interaction is given by,
\begin{align}
    \mathcal{L}_{\rm eff} \supset 
     G_{\rm eff}(\bar{\nu}\nu)(\bar{\nu}\nu), 
\end{align}
where $ G_{\rm eff}$ is the dimensionful effective coupling. Using the Planck observation, a fit to the CMB data identifies two regimes, namely ``strongly interacting" (SI) and ``moderately interacting" (MI), for the effective coupling $G_{\rm eff}$~\cite{Kreisch:2019yzn, Park:2019ibn},
\begin{align}
\label{eq:GeffVal}
    G_{\rm eff} = \begin{cases}
    \left(4.73^{+0.37}_{-0.61}~\text{MeV}\right)^{-2}~~\text{(SI)}\\
    \left(89.12^{+170.89}_{-60.94}~\text{MeV}\right)^{-2}~~\text{(MI)}
    \end{cases}
    \approx \begin{cases}
    \left(5~\text{MeV}\right)^{-2}~~\text{(SI)}\\
    \left(100~\text{MeV}\right)^{-2}~~\text{(MI)}
    \end{cases}.
\end{align}
Note that this $G_{\rm eff}$ is much larger than the Fermi constant $G_{\rm F}$.
It is also shown in~\cite{Kreisch:2019yzn, Park:2019ibn} that the SI case is preferable in ameliorating the $H_0$ tension as well as remaining consistent with the local astronomical observations. 
%

%
In our case, the self-interaction of neutrinos are mediated by the additional $Z^{\prime}$ arising due to the hidden $U(1)_H$ gauge symmetry. It is to be noted that there are other scalars in the model but since in our model neutrinos do not self-interact via scalar fields, they will have no effect in the explanation of the  Hubble tension.
Clearly, the effective coupling $G_{\rm eff}$ can be represented in terms of the present model parameters as,
\begin{align}
\label{eq:GeffModel}
G_{\rm eff} = \frac{g_H^2 \tilde{\epsilon}^4}{m_{Z^{\prime}}^2},  
\end{align}
where $\Tilde{\epsilon}$ represents the mixing between the neutral fermions as defined in the last part of Sec.~\ref{sbsc:gaugeSec}.
%
It has been shown in~\cite{Berbig:2020wve} that to maintain BBN and other constraints, $g_H\tilde{\epsilon}^2 \in [2\times 10^{-7}, 5\times 10^{-6}]$ 
%
%
and this, using Eqs.~\eqref{eq:GeffVal} and~\eqref{eq:GeffModel}, translates to the constraint on the $Z^{\prime}$ mass as,
\begin{align}
 m_{Z^{\prime}}\in \begin{cases}
     [1,25]~\text{eV~~(SI)}\\
     [20,500]~\text{eV~~(MI)}
 \end{cases}.   
 \label{eq:mzpRange}
\end{align}
Using Eq.~\eqref{eq:mzmzpMass} one can set VEVs $v_{\varphi}$, $v_{\varphi^{\prime}}$, and the dark sector gauge coupling $g_H$ to obtain $m_{Z^{\prime}}$ in the above-mentioned ranges. 
Below, we show two benchmark points of \textit{(i)} $(g_H\tilde \epsilon^2,m_{Z'})=(2\times10^{-7},1~{\rm eV})$, and \textit{(ii)} $(g_H\tilde \epsilon^2,m_{Z'})=(5\times10^{-6},25~{\rm eV})$ satisfying all the constraints that we have discussed before,
\begin{align}
(i) \quad &
\tilde\epsilon=0.0144,\\
(ii) \quad &
\tilde\epsilon=0.00288,
\end{align}
where we fixed $v_{\varphi}=v_{\varphi'}= 10^{-2}{\rm GeV},\ 
\rho=10^{-9}$ and then the following values are commonly obtained 
$\Delta= 3.70\times10^{-6}\ {\rm GeV},\ M_{Z'}= 7.59\times10^{-7}\ {\rm GeV},\  g_H=0.0242$.

\section{Summary and Conclusions}
\label{sec:sumConcl}
We have proposed a radiative seesaw model in a hidden gauge $U(1)$ symmetry.
In order to have anomaly cancellations, we need to introduce several new fermions that contribute to $(g-2)_\mu$ as well as the neutrino oscillation data. We have also considered a fermionic DM candidate that correlates
with $(g-2)_\mu$ and neutrino mass matrix at the same time.
 We have demonstrated allowed regions in our input parameters satisfying several constraints.
 Finally, we have briefly discussed the Hubble tension via a lighter hidden gauge boson. 
In order to achieve it, we have found that we need to have nonzero VEV of $\eta$ to mix with neutral fermions
that leads to tree-level neutrino mass matrix. But, we have confirmed that this mass matrix is negligible compared to the radiative seesaw model by choosing small VEV of $\eta$. In the process we can generate requisite neutrino self-interaction mediated by the additional gauge boson to address the Hubble tension.
 

\vspace{1cm}
\noindent
\textbf{Acknowledgments}

\vspace{0.25cm}
\noindent
{\it
UKD acknowledges the hospitality of APCTP Pohang where this work was initiated.
This research of HO was supported by an appointment to the JRG Program at the APCTP through the Science and Technology Promotion Fund and Lottery Fund of the Korean Government. This was also supported by the Korean Local Governments - Gyeongsangbuk-do Province and Pohang City. HO is sincerely grateful for the KIAS membership.}

\bibliography{MA4_emug2.bib}
\end{document}